\documentclass[wssquare]{ws-procs961x669} 
\usepackage[colorlinks=true]{hyperref}
\hypersetup{
    bookmarks=true,         
    unicode=false,          
    pdftoolbar=true,        
    pdfmenubar=true,        
    pdffitwindow=false,     
    pdfstartview={FitH},    
    pdftitle={S},    
    pdfauthor={S. Sachdev},     
    pdfsubject={},   
    pdfcreator={},   
    pdfproducer={}, 
    pdfkeywords={} {} {}, 
    pdfnewwindow=true,      
    colorlinks=true,       
    linkcolor=magenta, 
    citecolor=blue,        
    filecolor=magenta,      
    urlcolor=blue           
}

\begin{document}
\title{Quantum spin glasses and Sachdev-Ye-Kitaev models}

\author{Subir Sachdev$^*$}

\address{Department of Physics, Harvard University\\
Cambridge MA 02138, USA\\
$^*$E-mail: sachdev@g.harvard.edu\\
Web: sachdev.physics.harvard.edu}

\begin{abstract}
A brief survey of some random quantum models with infinite-range couplings is presented, ranging from the quantum Ising model to the Sachdev-Ye-Kitaev model.

The Sachdev-Ye-Kitaev model was the first to realize an extensive zero temperature entropy without requiring an exponentially large ground state degeneracy. This phenomenon is closely linked to the absence of a particle-like interpretation of its low energy spectrum—its spectral functions are not those of bosons or fermions but are instead “Planckian”, meaning they are universal functions of energy/temperature. A remarkable consequence of these properties is that the SYK model provides an effective low energy theory of non-supersymmetric charged or rotating black holes in 3+1 dimensions, leading to new results on the density of many-body quantum states of such black holes. 

For applications to non-quasiparticle metallic states of quantum materials, an extension of the SYK model, known as the two-dimensional Yukawa-Sachdev-Ye-Kitaev model, is required. The 2dYSYK model describes quantum phase transitions in metals with spatial inhomogeneity in the position of the quantum critical point. This extension has led to a universal theory of the strange metal state observed in numerous correlated electron compounds, including copper-oxide based high temperature superconductors.
\\
\begin{center}
{\tt Rapporteur presentation at the 29th Solvay conference,\\ {\it The Structure and Dynamics of Disordered Systems},\\ October 21, 2023, Brussels}\\
\href{https://arxiv.org/abs/2402.17824}{arXiv:2402.17824}.
\end{center}
\end{abstract}


\bodymatter

\section{Introduction}
\label{sec:intro}

Quantum effects on infinite-range models of spin glasses have been studied for many decades. The early work \cite{Yamamoto_1987,Kopec89,Ray89,Usadel90,Huse93,YSR93,RSY95,Grempel98, Kennett01,Rozenberg01,Muller12,Mukherjee15,Mukherjee18,Young17,Kiss23,Maria23,Lang:2024ekr} was focused primarily on adding a transverse Zeeman field $g$ to the Sherrington-Kirkpatrick model of an Ising spin glass \cite{SK75}, and related models. The spin operator along the transverse field direction does not commute with the spin operator in the Ising interaction, and so quantum effects must be including in determining its phase diagram. In these works, it was generally assumed, and verified by computations, that the effects of quantum fluctuations have many similarities to the effects of thermal fluctuations. Key features of the quantum Ising models are:
\begin{itemize}
\item
At large $g$, there is a unique `trivial' quantum paramagnetic ground state at large $g$, and this is smoothly connected to the thermal paramagnet at high temperature ($T$).
\item Within the spin glass state found at small $g$ and small $T$, the structure of the replica symmetry breaking in the quantum model is the same as that found in the classical model by Parisi \cite{Parisi79}. The main quantum effect is an overall renormalization of the Edwards-Anderson spin glass order parameter \cite{HertzBook}.
\item Quantum effects are innocuous in the long-time aging dynamics within the spin glass phase (apart from overall renormalizations), and the same equations apply in the classical and thermal models \cite{Sompolinsky1,Sompolinsky2,Sompolinsky3,Kurchan94,Kurchan95,Cugliandolo_Lozano,Biroli02,Schiro20}. These equations have an emergent time reparameterization invariance, but the solutions spontaneously break time translational symmetry and are not conformally invariant. 
\end{itemize}

In the meantime, studies of quantum phase transitions in spin systems without disorder, motivated by the physics of the cuprates, showed that there were strong differences between the effects of  thermal and quantum fluctuations. These differences appeared most prominently in situations where a `trivial' quantum ground state ({\it i.e.\/} a state smoothly connected to a site-product state) was prohibited at any coupling as a consequence of what are now often called `Lieb-Schultz-Mattis anomalies'. As an example, the spin $S=1/2$ square lattice antiferromagnet with arbitrary interactions which preserve full spin rotation symmetry cannot have a trivial ground state: it must either break a spin rotation or lattice symmetry, or have non-trivial degeneracies on a torus associated with the presence of fractionalized anyonic excitations. 

Motivated by my study with N. Read of examples of such anomalous effects of quantum fluctuations on the square lattice \cite{NRSS89,NRSS91}, I decided to examine an infinite-range random quantum spin model for which a trivial ground state was prohibited \cite{SY92}. This was the generalization of the Sherrington-Kirkpatrick model to a quantum Heisenberg model with SU($M$) spin rotation symmetry and spin $S$ (for general $M$, $S$ refers to the size of the SU($M$) representation), a variant of which is now called the Sachdev-Ye-Kitaev model \cite{kitaev2015talk,Maldacena:2016hyu}. 
For large $M$ in the Heisenberg model, and for the SYK model, a novel quantum-critical state is obtained \cite{SY92}, with many interesting properties. This quantum critical state has no analog in the Ising model in a transverse field. Key features of the SYK critical state are:
\begin{itemize}
\item There are no quasiparticle excitations. This distinguishes the SYK model from earlier solvable models which are `integrable'. 
\item There is an extensive zero temperature entropy, with the $T \rightarrow 0$ limit taken after the $N\rightarrow \infty$ limit ($N$ is the system size) \cite{GPS00,GPS01}. The SYK model was the first model to have this feature {\it without} an exponentially large ground state degeneracy. Earlier models, such as Pauling's ice model \cite{PaulingIce}, require fine tuning by an infinite number of parameters, and do have an exponentially large ground state degeneracy and an energy gap above the ground state. The SYK model realizes the extensive $T \rightarrow 0$ entropy by a level spacing exponentially small in system size above the ground state. These properties are crucial for the black hole connection.
\item The SYK saddle-point correlators have an emergent conformal SL(2, $R$) symmetry \cite{PG98}. This is also important for the black hole mapping, and has no analog in the classical spin glass model. The conformal symmetry along with the strongly coupled nature implies that dynamics is `Planckian' {\it i.e.\/} thermalization and relaxation times are of order $\hbar/k_B T$, a feature shared with both strange metals and black holes.
\item There is an emergent time reparameterization invariance in the SYK critical state \cite{kitaev2015talk,Maldacena:2016hyu}. But there is no aging glassy dynamics and time translational symmetry is now preserved.
\end{itemize}

Remarkably, the SYK critical state has found numerous physical applications, some far removed from the original motivation for its study:
\begin{itemize}
\item I proposed in 2010 \cite{SS10} that `certain mean-field gapless spin liquids' are quantum matter states without quasiparticle excitations realizing the low energy quantum physics of charged black holes. With `mean-field gapless spin liquids' I was referring to what can now be called the SYK critical state. I argued for a correspondence between the SYK model and charged black holes at the semiclassical level, based upon a remarkable match between the behaviors of the entropy and correlation functions. In 2015, Kitaev \cite{kitaev2015talk} showed that the correspondence held at the fully quantum level.  This connection has undergone rapid development in recent years, and has led to an understanding of the generic universal structure of the low-energy density of states of non-supersymmetric charged and rotating black holes in $D \geq 4$ spacetime dimensions \cite{Banerjee:2010qc,Sen12,Nayak:2018qej,Moitra18,Sachdev19,Iliesiu:2020qvm,Iliesiu:2022onk,Turiaci_Review,Moitra:2019bub,Kapec24a,Kolanowski:2024zrq,Kapec24b}. 
\item A variant of the SYK model couples fermions and bosons with a random Yukawa coupling, and is sometimes called the Yukawa-SYK model \cite{Fu16,Murugan:2017eto,Patel:2018zpy,Marcus:2018tsr,Wang:2019bpd,Ilya1,Wang:2020dtj,KimAltman20,WangMeng21,Schmalian2,Schmalian3}. An extension of the Yukawa-SYK model to finite spatial dimension $d=2$, the 2dYSYK model, has led to a realistic universal model of the strange metal state of correlated electron systems \cite{Altman1,Patel1,Maria22,Patel2,Guo2022,Schmalian1,PatelLunts,Li:2024kxr,HardyPatel24,AAPQMC}.
\end{itemize}
These applications will be reviewed in Sections~\ref{sec:bh} and \ref{sec:sm}.

\section{Sherrington-Kirkpatrick model}
\label{sec2}

The degrees of freedom of the classical Sherrington-Kirkpatrick model \cite{SK75} are Ising spins $Z_i = \pm 1$ on a set of sites $i =1 \ldots N$. These spins are coupled together by the Hamiltonian
\begin{align}
H_1  &= \frac{1}{2\sqrt{N}}\sum_{i,j=1}^N J_{ij} Z_i \, Z_j \label {e1} \\
\overline{J_{ij}} = 0, \quad \overline{J_{ij}^2} = J^2 &, \quad \mbox{different $J_{ij}$ uncorrelated.} \nonumber
 \end{align}
where $J_{ij}$ are independent random variables with zero mean and variance $J^2$. We are interested in the properties of thermal partition function
\begin{align}
\mathcal{Z}_1 & = \sum_{Z_i= \pm 1} e^{-H_1/T}\,,
\end{align}
at a temperature $T$. 
\begin{figure}[h]
\begin{center}
\includegraphics[width=1in]{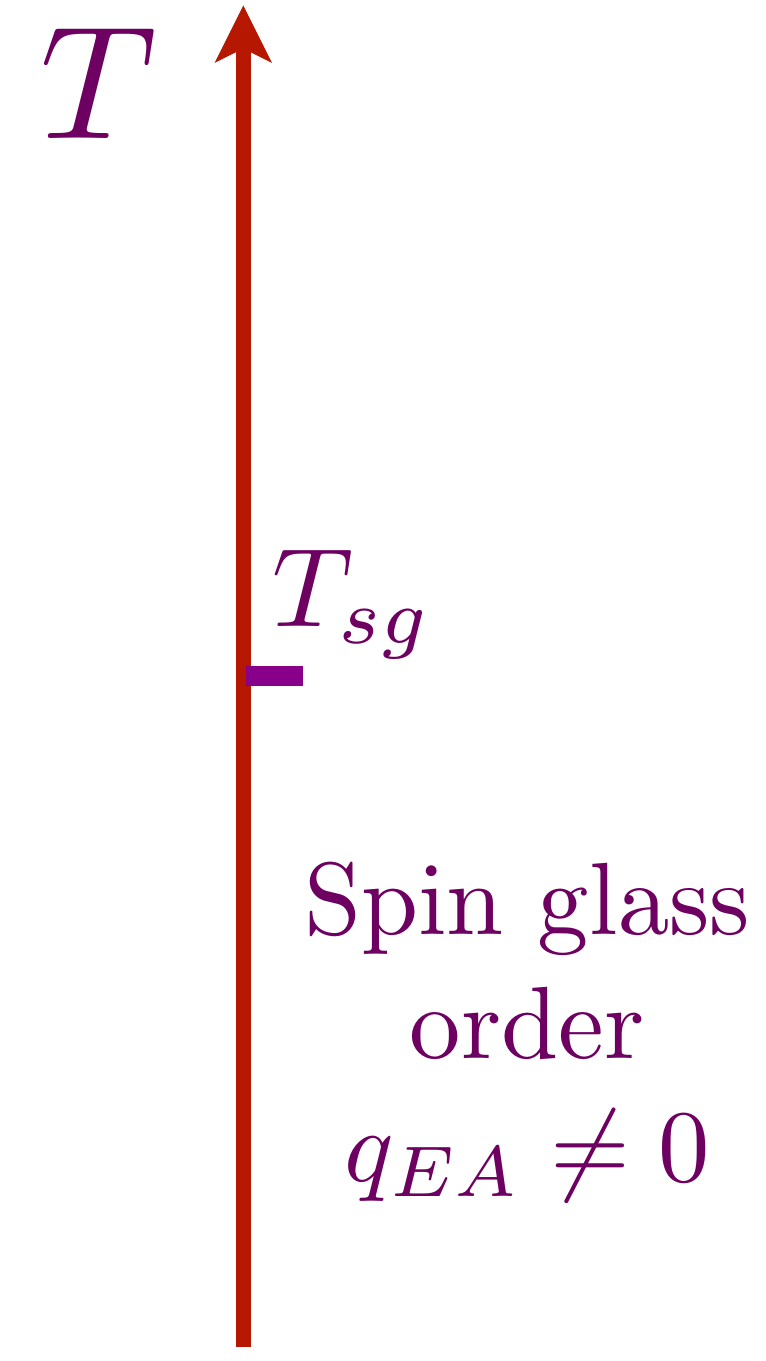}
\end{center}
\caption{Phase diagram of the Sherrington-Kirkpatrick model in (\ref{e1}).}
\label{fig1}
\end{figure}
The well-known phase diagram is sketched in Fig.~\ref{fig1}. There is a single phase transition at $T=T_{sg}$. For $T>T_{sg}$, we have the high temperature paramagnet. For $T<T_{sg}$, we obtain the spin glass state in which the 
Edwards-Anderson order parameter 
\begin{align}
q_{EA} = \overline{ \langle Z_i \rangle^2}
\end{align}
is non-zero (the overline represents the average over the ensemble of $J_{ij}$). The full structure of the spin glass state is characterized by Parisi's replica symmetry breaking (RSB) ansatz \cite{Parisi79}.

\section{Quantum Ising model}
\label{sec:Ising}

We obtain the quantum Ising model from (\ref{e1}) by adding a Zeeman field coupling in the `$x$' direction
\begin{align}
 H_2  & = \frac{1}{2\sqrt{N}}\sum_{i,j=1}^N J_{ij} Z_i Z_j - g \sum_{i=1}^N X_i \label{e2} \\
\mathcal{Z}_2 & = \mbox{Tr}\, e^{-H_2/T}\,. \nonumber 
\end{align}
Now $Z_i$ is a $z$-Pauli matrix acting on the 2 states of the classical model, and $X_i$ is the $x$-Pauli matrix on each site.
Quite apart from its theoretical interest, models closely related to $H_2$ describe numerous recent quantum devices \cite{Ebadi22,King23,Maciej23}.

\begin{figure}[h]
\begin{center}
\includegraphics[width=4in]{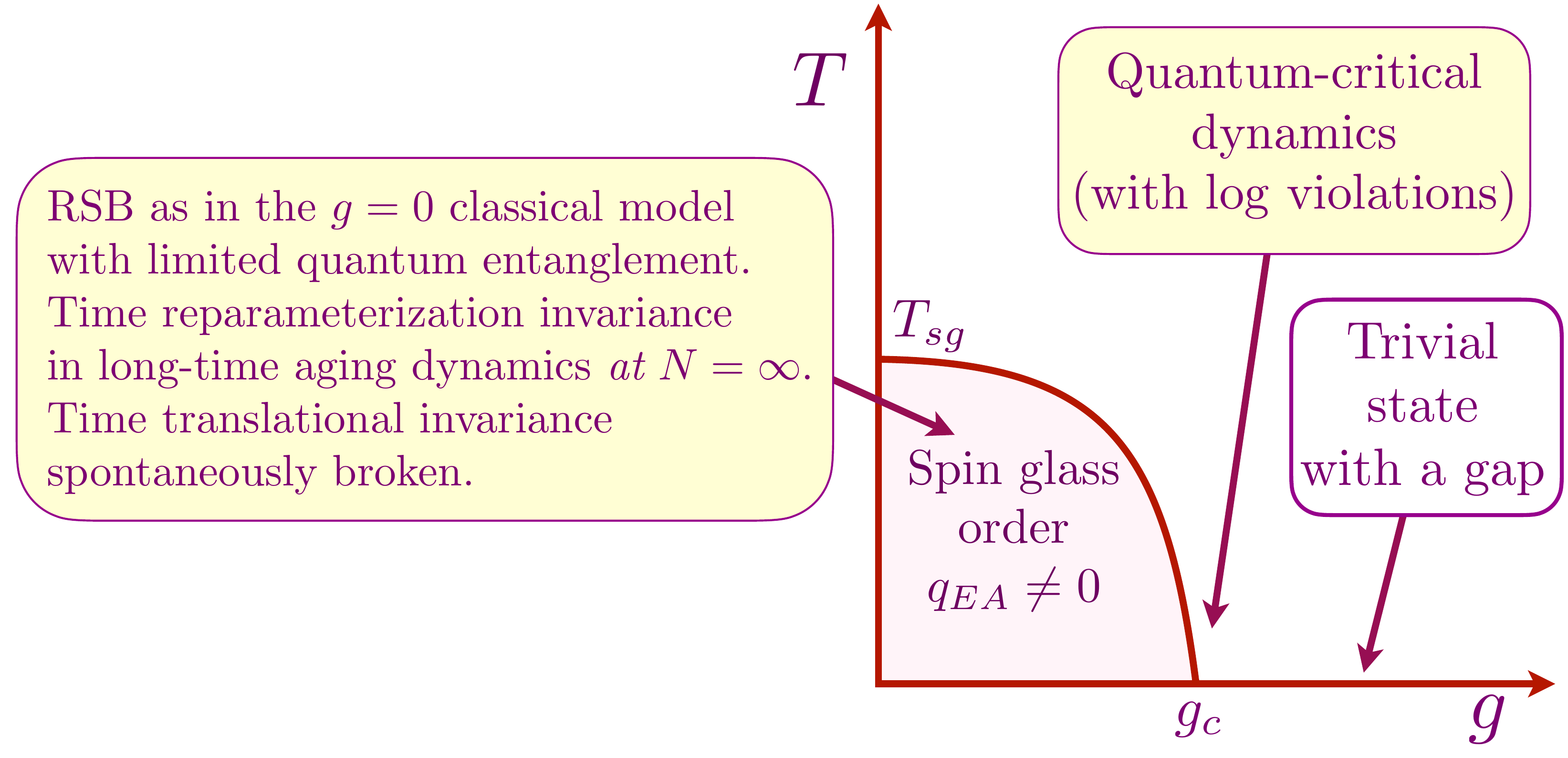}
\end{center}
\caption{Phase diagram of the quantum Ising model in (\ref{e2}).}
\label{fig2}
\end{figure}

We sketch the phase diagram of $H_2$ in Fig.~\ref{fig2}.
There is a phase with spin glass order at small $g$ and $T$, whose structure is similar to that of the classical model in Fig.~\ref{fig1}. A theory of a quantum model necessarily involves dynamics, and so now we can define the spin glass order parameter by the autocorrelation function on a single site;
at $T=0$ this is conveniently expressed in terms of the imaginary time ($\tau$) correlation
\begin{align}
q_{EA} = \lim_{\tau \rightarrow \infty} \overline{ \langle Z_i (\tau) Z_i (0) \rangle}\,.
\end{align}
We can move out of the spin glass phase either by raising temperature or by increasing $g$. In the latter case, we obtain a gapped paramagnetic ground state, smoothly connected to the trivial product state with all spins oriented in the $x$ direction. Near the quantum critical point at $g=g_c$, there is an interesting regime of quantum critical spin dynamics with a characteristic time of order $\hbar/(k_B T)$, with a logarithmic sensitivity to the magnitude of the exchange interactions $J$ \cite{YSR93,RSY95}. 

The real time quantum dynamics within the spin glass phase had mainly been studied for a spherical $p$-rotor quantum model \cite{Cugliandolo_Lozano,Biroli02}, but has recently been extended to the quantum Ising model \cite{Lang:2024ekr}. The $p$-rotor model has many similarities to the Ising model, but there is important difference that the $p$-rotor model has
only one-step replica symmetry breaking while the Ising model has full replica symmetry breaking, and this leads to some differences in their long-time dynamics. 
An important feature of the glassy dynamics in both cases is the emergence of time reparameterization invariance at long times: the $N=\infty$ equations of motion for the spin autocorrelator $\mathcal{C} (t_1, t_2)$  are invariant under the transformation \cite{Sompolinsky1,Sompolinsky2,Sompolinsky3,Kurchan94,Kurchan95}
\begin{align} 
 \mathcal{C} (t_1, t_2) \rightarrow
 \left[f'(t_1) f'(t_2) \right]^{\Delta} \mathcal{C} (f(t_1), C(f(t_2))
 \label{timepar1}
 \end{align}
 where $f(t)$ is a monotonic time reparameterization, and $\Delta$ is an exponent. However, despite this invariance, the solution to the full equations exhibits aging dynamics \cite{Kurchan94,Kurchan95,Cugliandolo_Lozano,Biroli02,Schiro20,Lang:2024ekr} in which time translational invariance is spontaneously broken. All these features are just as those for the classical model, and there are also connections between the aging dynamics and the structure of the replica symmetry breaking in the equilibrium solution. 

\section{Quantum Heisenberg model}
\label{sec:Heisenberg}

Next, we consider quantum models without a trivial paramagnetic ground state because of LSM anomalies. For this, we need a spin model in which the on-site states remain doubly degenerate in the absence of couplings between the sites. We consider the `Heisenberg' model with a global SU(2) spin rotation symmetry \cite{BrayMoore}
\begin{align}
 H_3  & = \frac{1}{2\sqrt{N}}\sum_{i,j=1}^N J_{ij} \left(X_i X_j + Y_i Y_j + Z_i Z_j \right)\,, \label{e3}
 \end{align}
where $X_i$, $Y_i$, $Z_i$ are the 3 Pauli operators on each site. Such a model is connected to recent studies with ultracold atoms \cite{Monika22,Esslinger23,Lev23}. It is also useful to generalize this model to SU($M$) symmetry, and consider the phase diagram for general $M$ \cite{SY92}.

\begin{figure}[h]
\begin{center}
\includegraphics[width=4in]{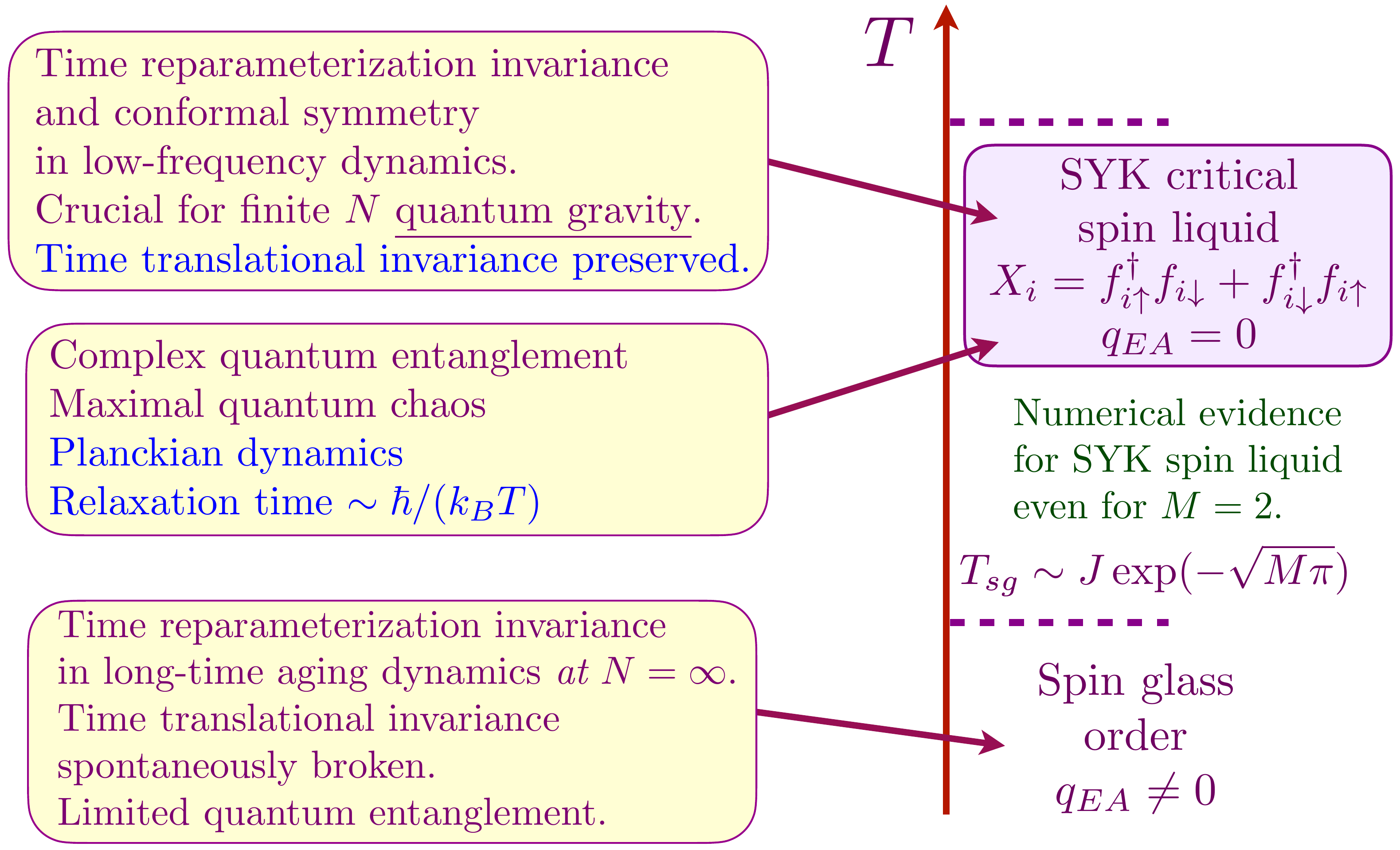}
\end{center}
\caption{Phase diagram of the quantum Heisenberg model in (\ref{e3}), generalized to SU($M$) symmetry.}
\label{fig3}
\end{figure}
Fig.~\ref{fig3} sketches the basic features of the phase diagram of the SU($M$) spin models, as deduced from various studies \cite{Chowdhury:2021qpy,SY92,PG98,GPS00,GPS01,Biroli02,MJR02,MJR03,Henry21,Dumi22,Christos:2021wno,GP23}. The low $T$ spin glass state is similar to that of the quantum Ising models, but a recent study has noted some interesting differences \cite{GP23}. 

The remarkable new feature of Fig.~\ref{fig3} is the `SYK spin liquid' regime which emerges in the intermediate temperature range $J e^{-\sqrt{\pi M}} < T < J$; there is evidence in Ref.~\cite{Henry21} that such a regime is identifiable even for $M=2$. This regime is best understood as one in which there is complex quantum entanglement leading to {\it fractionalization\/} of the spins. Formally, we can rewrite $H_3$ exactly as a model of interacting fermions $f_{i \alpha}$, $\alpha = \uparrow, \downarrow$ by using the decomposition and constraints
\begin{align}
(X_i, Y_i, Z_i) = \frac{1}{2} f_{i \alpha}^\dagger {\bm \sigma}_{\alpha\beta} f_{i \beta}^{\phantom \dagger}\quad , \quad f_{i \alpha}^\dagger f_{i \alpha}^{\phantom\dagger} = 1 ~\forall i\,, \label{Sparton}
\end{align}
where ${\bm \sigma}$ are the Pauli matrices. The SYK spin liquid appears when the $f_{i \alpha}$ fermions form the gapless critical state without quasiparticle excitations to be discussed in Section~\ref{sec:syk}. The properties of this state are summarized in Fig.~\ref{fig3}, and will be discussed further in Section~\ref{sec:syk}. There is an emergent time reparameterization symmetry here, but it differs from that in the spin glass phase in important respects: ({\it i\/}) time translational symmetry is preserved, ({\it ii\/}) the large $N$ saddle-point has conformal symmetry. Both these features are crucial in the connection to quantum gravity discussed in Section~\ref{sec:bh}.

There have been recent proposals \cite{Hanada23,Swingle23} that the SYK spin liquid state is present down to zero temperature in models of SU(2) spins with 4 or higher spin interactions, but that has been questioned by a rigorous analysis of related issues \cite{King24}. 

\section{Free fermions}
\label{sec:free}

\begin{figure}
\begin{center}
\includegraphics[width=3.5in]{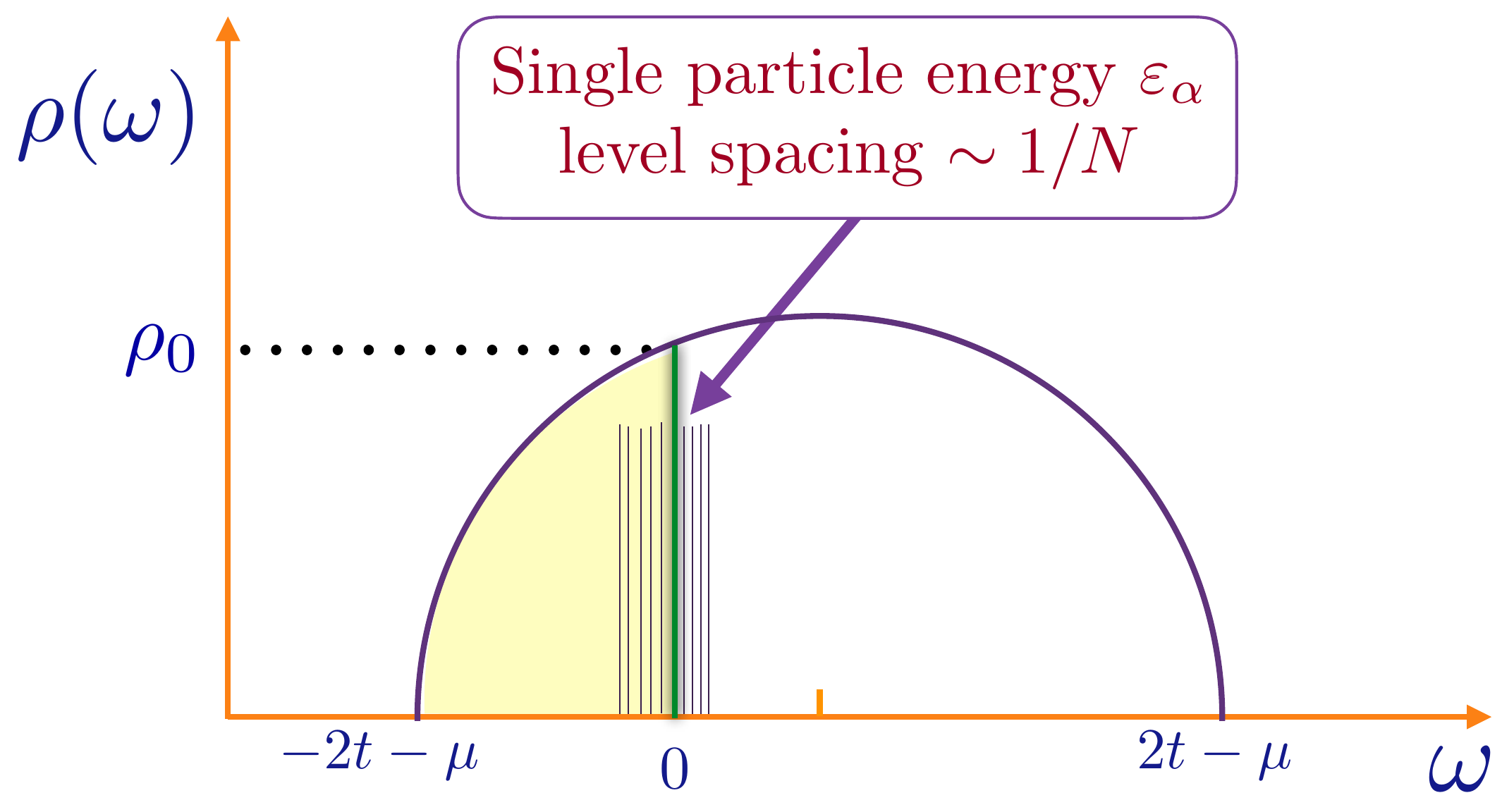}
\end{center}
\caption{Single particle density of states of $H_4$.}
\label{fig4}
\end{figure}

Before moving to the SYK model, it is useful to recall a few properties of a simple problem of free fermions with random hopping.
We consider fermions $c_i$ with hopping amplitudes $t_{ij}$ which are independent random numbers, with Hamiltonian 
\begin{align}
H_4 = & \frac{1}{\sqrt{N}} \sum_{i,j=1}^N  t_{ij} c_{i}^\dagger  c_{j} - \mu \sum_i c_i^\dagger c_i \\
\overline{t_{ij}} = 0, \quad &\overline{|t_{ij}|^2} = t^2 , \quad \mbox{different $t_{ij}$ uncorrelated.} \nonumber
\end{align}
This model is easily solved by determining the eigenvalues $\varepsilon_\alpha$ ($\alpha = 1 \ldots N$) of the random matrix $t_{ij} - \mu \delta_{ij}$. The distribution of these eigenvalues obeys Wigner-Dyson statistics: they are roughly equally spaced with a spacing of order $1/N$, and realize a semi-circular density of states, as shown in Fig.~\ref{fig4}.
Here $\rho(\omega)$ is the {\it single\/} particle density of states
\begin{align}
\rho (\omega) = \frac{1}{N} \sum_{\alpha} \delta (\omega - \varepsilon_{\alpha})\,.
\end{align}

The many-particle ground state of $H_4$ is obtained by occupying all states with $\varepsilon_\alpha < 0$. The familiar 
Sommerfeld theory yields the linear-$T$ dependence of the entropy at low $T$:
\begin{align}
 S(T \rightarrow 0)  = N \gamma T \quad, \quad
\gamma  = \frac{\pi^2}{3} \rho_0\,, \label{ST}
\end{align}
where $\rho_0 = \rho (0)$.

\begin{figure}
\begin{center}
\includegraphics[width=4in]{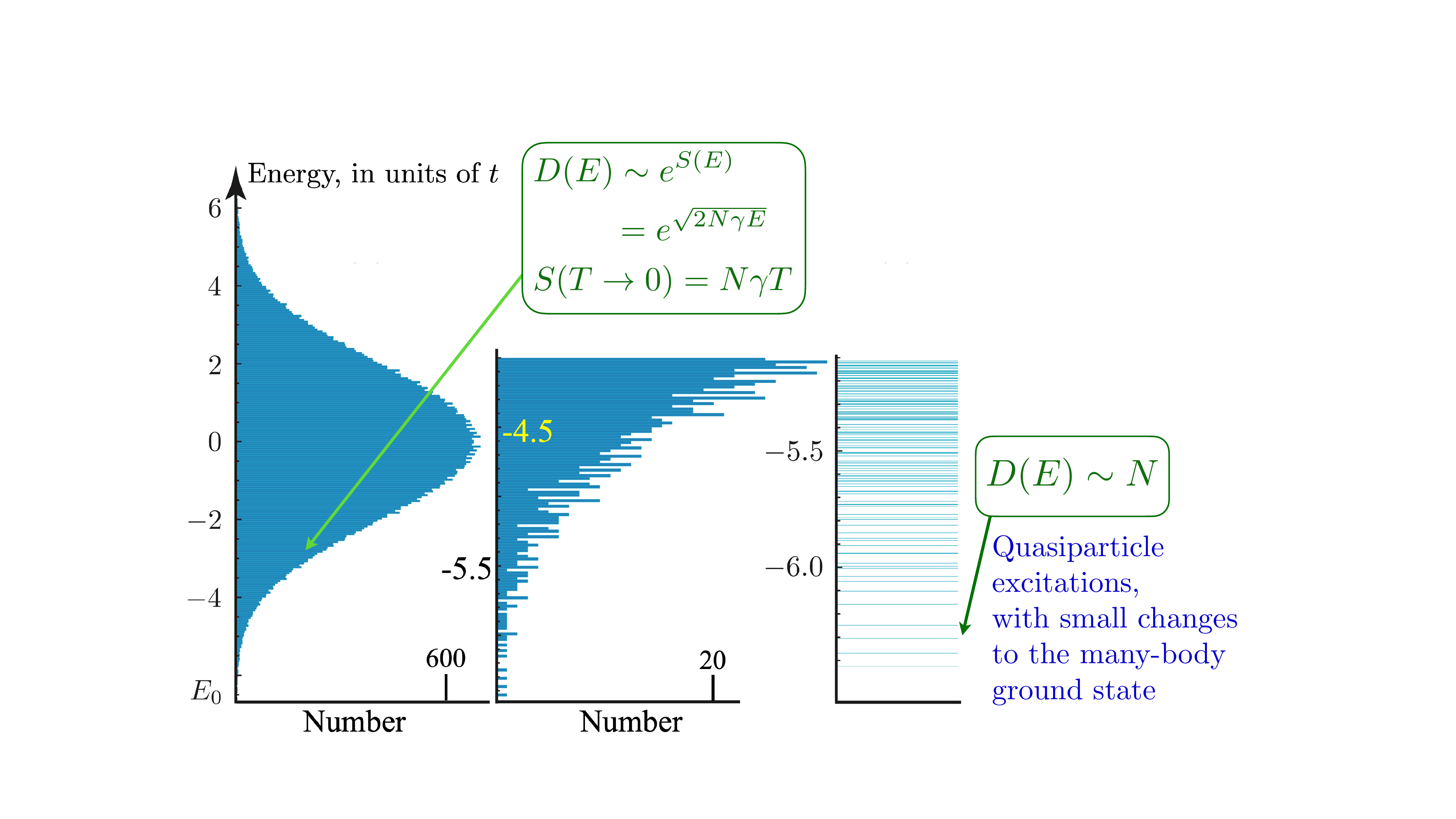}
\end{center}
\caption{The many-body density of states of $H_4$. Adapted from numerics by G.~Tarnopolsky and Ref.~\cite{Chowdhury:2021qpy}.}
\label{fig5}
\end{figure}

It is useful to map these results to a somewhat unfamiliar many-body perspective, as that will be useful in the comparison with the SYK model. In the grand-canonical ensemble, there are $2^N$ eigenstates $E_0 + E_a$, $a = 1\ldots 2^N$ given by
\begin{align}
E_0 + E_a = \sum_{\alpha} n_{\alpha} \varepsilon_{\alpha}
\end{align}
with $n_\alpha = 0, 1$, the occupation of the state $\alpha$. Here $E_0$ is the many-body ground state energy, chosen so that the smallest $E_a = 0$. We now study the behavior of the many-boday density of states
\begin{align}
D(E) = \sum_{a=1}^{2^N} \delta (E - E_a)
\end{align}
This has not been computed analytically for all $E$ and we show numerical results for small $N$ in Fig.~\ref{fig5}.

In the regime where $E$ is extensive (of order $N$), we can compute $D(E)$ by Boltzmann's relation
\begin{align}
D(E) \sim \exp( S(E)) \label{f1}
\end{align}
where $S(E)$ is the entropy in the micro-canonical ensemble. This entropy can be obtained for $E \ll Nt$ (but of order $N$) from the grand-canonical result for $T \ll t$ in (\ref{ST})
\begin{align}
S(E) = \sqrt{2 N \gamma E}\quad , \quad  D(E) \sim \exp (\sqrt{2 N \gamma E} ) \,. \label{f2}
\end{align}
So
\begin{align}
\ln D(E) \sim N \quad, \quad E \sim N\,, 
\end{align}
as expected in the thermodynamic limit. 

Let us also consider the case when $E$ approaches zero, and is of order $1/N$. Then the many-body eigenstates are obtained by adding or removing a small number of particles near the Fermi level. As the single-particle level spacing is $\sim 1/N$ (see Fig.~\ref{fig4}), we estimate
\begin{align}
D(E) \sim N \quad , \quad E \sim 1/N\,. \label{f3}
\end{align}

\section{SYK model}
\label{sec:syk}

We now allow the fermions in the $\varepsilon_\alpha$ eigenstates of Section~\ref{sec:free} to interact with each other. Given the random nature of these single-particle eigenstates, the matrix elements for interaction-induced transitions between these states will also be random. The complex SYK model \cite{Sachdev15,GKST} is obtained by keeping only the random interactions in the Hamiltonian acting on the $2^N$ many-body states of Section~\ref{sec:free}:
\begin{align}
&H_{5} = \frac{1}{(2 N)^{3/2}} \sum_{\alpha,\beta,\gamma,\delta=1}^N U_{\alpha\beta;\gamma\delta} \, c_\alpha^\dagger c_\beta^\dagger c_\gamma^{\vphantom \dagger} c_\delta^{\vphantom \dagger} 
-\mu \sum_{\alpha} c_\alpha^\dagger c_\alpha^{\vphantom \dagger} \label{e5} \\
\overline{U_{\alpha\beta;\gamma\delta}} & = 0, \quad \overline{|U_{\alpha\beta;\gamma\delta}|^2} = U^2 , \quad \mbox{different $U_{\alpha\beta;\gamma\delta}$ uncorrelated.} \nonumber
\end{align}
It is also useful to define the charge density $\mathcal{Q}$
\begin{align}
\mathcal{Q} = \frac{1}{N} \sum_\alpha c_\alpha^\dagger c_\alpha^{\vphantom \dagger} \, ; \quad
[{H}_5, \mathcal{Q}] = 0\, ; \quad  0 \leq \mathcal{Q} \leq 1\,.
\end{align}
There are now several experimental proposals or realizations of systems with SYK-like Hamiltonians 
\cite{chew2017, pikulin2017,Garcia-Alvarez:2016wem,Babbush:2018mlj,danshita2017,Tigran21,Uhrich:2023ddx,chen2018, brzezinska2022,jafferis2022,luo2019,Laurel23,Sonner24}. A model closely related to $H_5$ is obtained when the fermonic spin representation in (\ref{Sparton}) is inserted in the quantum Heisenberg spin model in (\ref{e3}) \cite{SY92}. 

\begin{figure}
\begin{center}
\includegraphics[width=4in]{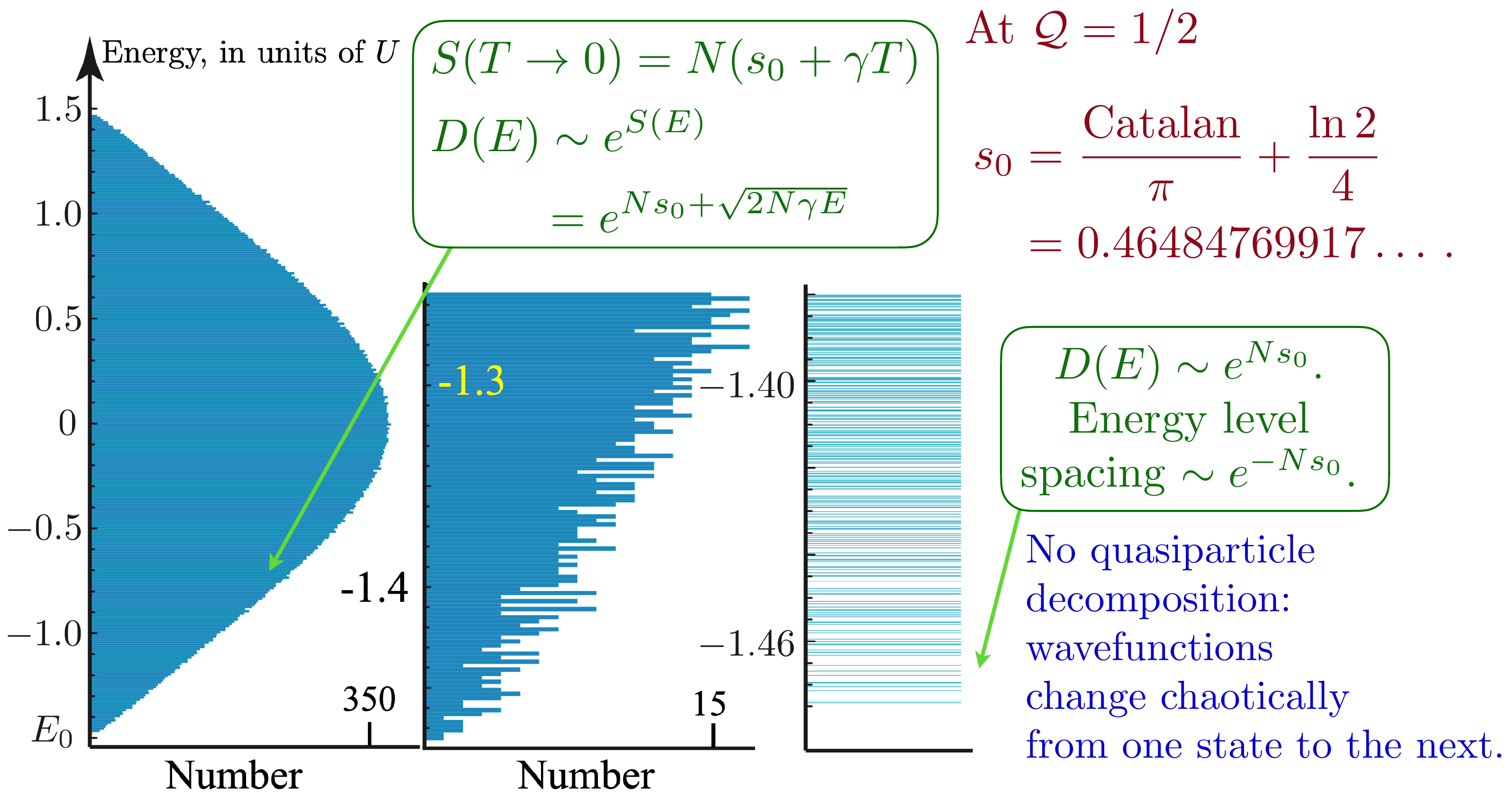}
\end{center}
\caption{The many-body density of states of $H_{5}$. Adapted from numerics by G.~Tarnopolsky and Ref.~\cite{Chowdhury:2021qpy}.
Compare to Fig~\ref{fig5}, where $D(E) \sim N$ at low energy.}
\label{fig6}
\end{figure}

Numerical results for the spectrum of $H_5$ appear in Fig.~\ref{fig6}. 
The important difference from the free fermion results in Fig.~\ref{fig5} is the exponentially larger density of states at low energies, as is discussed further below.

There is no glassy behavior in the ground state of the SYK model \cite{Gur-Ari:2018okm}, and recent rigorous considerations \cite{King24} have shown a clear distinction between the very low energy states of quantum spin glass models (including those described in Sections~\ref{sec:Ising} and \ref{sec:Heisenberg} (for $M$ finite)) and the SYK model.
Consequently, we can proceed by examining only the replica diagonal sector of
the averaged partition function of $H_5$. This can be written as path integral over
the bilocal fermion field
$G(\tau_1, \tau_2) \sim ({1}/{N}) \sum_\alpha c_\alpha (\tau_1) c_{\alpha}^\dagger (\tau_2)$ \cite{GPS01,Maldacena:2016hyu,kitaevsuh,Sachdev15}
\begin{align}
\overline{\mathcal{Z}}_5 = \int \mathcal{D} G(\tau_1, \tau_2) \exp \left( - N S_{\rm eff} [ G] \right) \label{e5a}
\end{align}
The large $N$ saddle point equation $\delta S_{\rm eff}/\delta G = 0$ yield a time-translational invariant solution for the fermion Green's function $G_s (\tau_1 - \tau_2)=\left\langle G(\tau_1,\tau_2) \right\rangle$ \cite{SY92}: 
\begin{align}
G_s (i\omega) = \frac{1}{i \omega + \mu - \Sigma_s (i\omega)} \quad &, \quad \Sigma_s (\tau) = -  U^2 G_s^2 (\tau) G_s(-\tau)\,.
\label{sy}
\end{align}
These equations also apply unchanged to the Green's function and self energy of the $f$ fermions in (\ref{Sparton}) for the Heisenberg spin model 
of Section~\ref{sec:Heisenberg} with SU($M \rightarrow \infty$) symmetry \cite{SY92}. 
The solution of (\ref{sy}) yields the basic properties of the SYK critical state.

A key feature is the emergence of a time reparameterization symmetry at low frequencies and temperatures much smaller than $U$, where the $i \omega + \mu$ term in (\ref{sy}) can be neglected.
Then the action $S_{\rm eff}[G]$ becomes invariant 
invariant under time reparametrization $f(\sigma)$ \cite{kitaev2015talk,Maldacena:2016hyu}:
\begin{align}
\tau &= f (\sigma) \nonumber \\
G(\tau_1 , \tau_2) &= \left[ f' (\sigma_1) f' (\sigma_2) \right]^{-1/4} \widetilde{G} (\sigma_1, \sigma_2) 
\label{timepar2} 
\end{align}
Note the similarity of (\ref{timepar2}) to the classical time reparameterization in (\ref{timepar1}). Another key feature of the SYK critical state is that the saddle-point Green's function has an invariance under SL(2, $R$) conformal transformations \cite{PG98}:
\begin{align}
G_s (\tau_1 - \tau_2 ) = \widetilde{G}_s (\sigma_1 - \sigma_2)~\text{for}~ \tau = \frac{a \sigma + b}{c  \sigma +d}\,,~ad-bc=1\,. 
\label{sl2r}
\end{align}
Combined with the strongly-coupled nature, this property implies that the dynamics is `Planckian' at $T>0$ with a relaxation time $\sim \hbar/(k_B T)$ \cite{PG98,Eberlein}.
There is no analog of (\ref{sl2r}) for the spin glass state reparameterization in (\ref{timepar1}). The symmetry (\ref{sl2r}) will be crucial for the black hole mapping, as we will see in Section~\ref{sec:bh}. The path integral in (\ref{e5a}) also has an emergent U(1) gauge symmetry, linked to the presence of the conserved charge $\mathcal{Q}$ \cite{Sachdev15,Davison17,GKST}.

The time reparameterization and gauge symmetry are hints that the low energy effective theory of the SYK model is a theory of gravity and electromagnetism. Indeed, the theory turns out to be just Einstein-Maxwell theory in the background of a charged black hole, as will be discussed further in Section~\ref{sec:bh}. Furthermore, the path integral in (\ref{e5a}) can be exactly evaluated for this effective theory \cite{kitaev2015talk,Maldacena:2016hyu,cotler16,Bagrets17,StanfordWitten,GKST}, leading to the following finite $N$ results for the low energy many-body density of states in the ensemble with a fixed $\mathcal{Q}$ \cite{GKST}:
\begin{align}
D(E, \mathcal{Q}) \sim \frac{1}{N} \exp (N s_0) \sinh \left( \sqrt{2 N \gamma E} \right)\,, \quad E \ll NU\,. \label{de}
\end{align}
The number $s_0$ is a universal function of $\mathcal{Q}$, computed in Ref.~\cite{GPS01}.
For extensive $E$, the $E$ dependence of (\ref{de}) is the same as (\ref{f2}) for the free fermion model, but with an exponentially large $E$-independent prefactor. 
\begin{align}
D(E, \mathcal{Q}) \sim \exp (N s_0) \exp(\sqrt{2 N \gamma E} ) \quad, \quad E \sim N\,.
\end{align}
However, for small $E$ there is a dramatic difference from the free fermion model, as indicated in Figs.~\ref{fig5} and \ref{fig6}, with 
\begin{align}
D(E, \mathcal{Q}) \sim \sqrt{\frac{E}{N}} \exp (Ns_0) \quad, \quad E \ll 1/N \,,
\end{align}
in contrast to (\ref{f3}).

The partition function of the SYK model is related to (\ref{de}) by
\begin{align}
\overline{\mathcal{Z}}_5 (T, \mathcal{Q}) = \int_{0}^\infty dE D(E, \mathcal{Q})e^{-E/T}\,. \label{f5}
\end{align}
For the canonical entropy of the SYK model at low $T$, (\ref{f5}) yields \cite{Maldacena:2016hyu,kitaevsuh,GKST}
\begin{align}
S(T, \mathcal{Q}) = N(s_0 + \gamma T) - \frac{3}{2} \ln \left(\frac{U}{T}\right) - \frac{1}{2} \ln N \,. \label{ST2}
\end{align}
Comparing to the free fermion result in (\ref{ST}), we notice an extensive entropy $Ns_0$ in the low $T$ limit \cite{GPS01}, and this will map to the $T=0$ Bekenstein-Hawking entropy of charged black holes in Section~\ref{sec:bh}. Note that the ground state is not degenerate, and this extensive entropy does not imply an exponentially large ground state degeneracy: the latter requires $T \rightarrow 0$ before $N \rightarrow \infty$, and here we are taking the opposite order of limits. 
The sub-extensive term in (\ref{ST2}) has a singular $-(3/2)\ln (1/T)$ dependence on temperature, and this implies that the expansion in (\ref{ST2}) breaks down at a temperature exponentially small in $N$. Given the exponentially small level spacing near the ground state (see Fig.~\ref{fig6}), this is reasonable because we have to account for the discreteness of the energy levels at such low $T$.

\section{Quantum black holes}
\label{sec:bh}

See Ref.~\cite{SSMEF} for a more detailed review of the topics discussed in this section. 

\begin{figure}
\begin{center}
\includegraphics[width=4.5in]{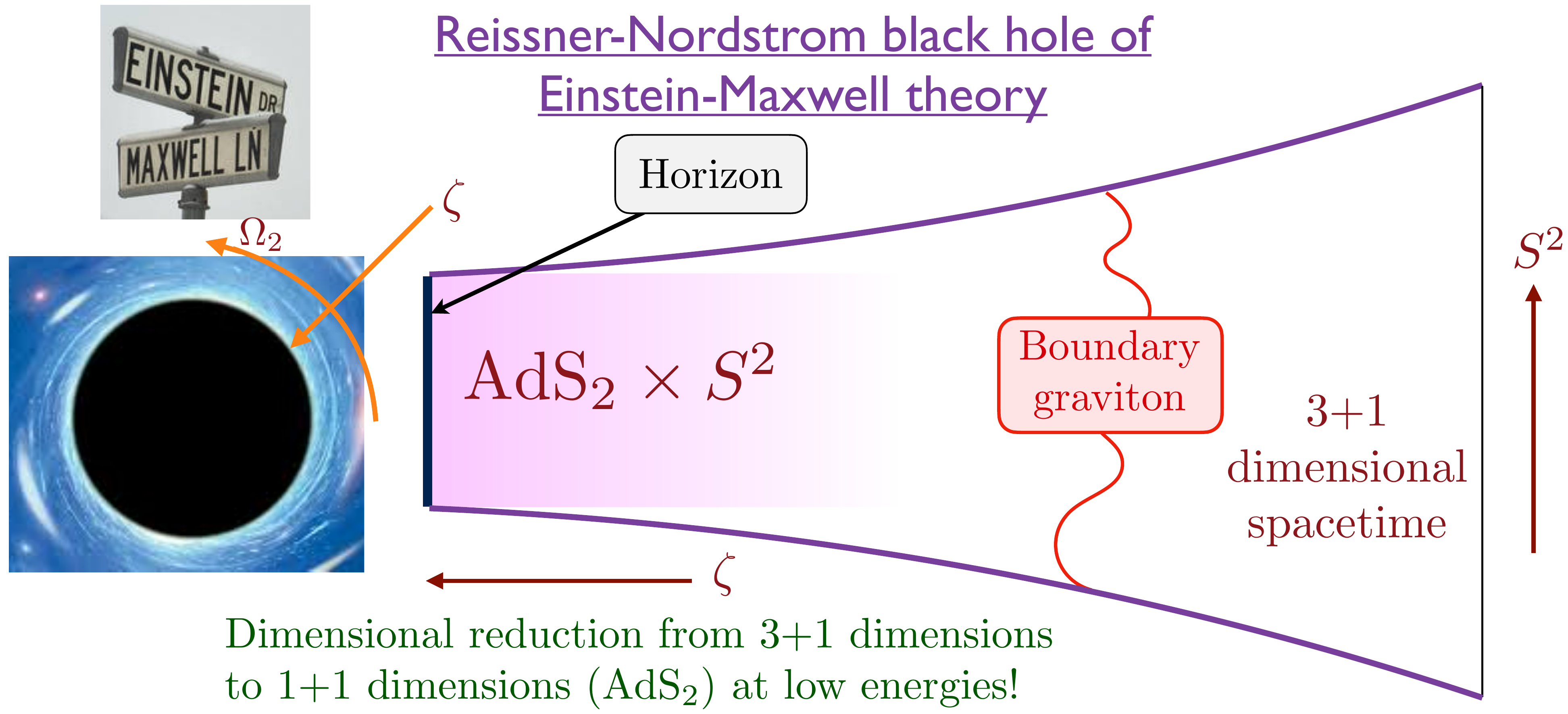}
\end{center}
\caption{Near-horizon AdS$_2$ geometry of charged black holes.}
\label{fig7}
\end{figure}

A key fact linking charged black holes to the SYK model is illustrated in Fig.~\ref{fig7}: the solution of the Einstein-Maxwell equations for a black hole with a net charge $\mathcal{Q}$ in spatial dimensions $d \geq 3$ yields a near-horizon metric with a AdS$_2$ factor \cite{Myers99,Cubrovic:2009ye,Liu2}. And the isometry of AdS$_2$ is SL(2, $R$) identical to the symmetry of the SYK saddle point in (\ref{sl2r}).
The AdS$_2$ metric in Euclidean time is
\begin{align}
ds^2 = \frac{d \tau^2 + d \zeta^2}{\zeta^2}\,, \label{appads2}
\end{align}
and this is invariant under the co-ordinate change
\begin{align}
\tau' + i \zeta' = \frac{a (\tau + i \zeta) + b}{c (\tau + i \zeta) + d}\,, \quad ad-bc =1\,.
\end{align}

Gibbons and Hawking \cite{Gibbons_Hawking} proposed to determine the quantum thermodynamics of a black hole by computing the path integral of Einstein-Maxwell theory in imaginary time outside the black hole horizon. The black hole metric in imaginary time has a cigar geometry which closes off at the horizon, and so this allowed them to sidestep the lack of knowledge of the black hole interior. 
They evaluated the partition function of a black hole with charge $\mathcal{Q}$ in an asymptotically 3+1 dimensional Minkowski spacetime
\begin{align}
\mathcal{Z}_6 (\mathcal{Q},T) =  \int \mathcal{D} g_{\mu\nu} \mathcal{D} A_\mu \exp \left( - \frac{1}{\hbar} ~ I^{(3+1)}_{\mathrm{Einstein~gravity}
+\mathrm{Maxwell~EM}} [g_{\mu\nu},A_\mu] \right)  \,,
\end{align}
where $g_{\mu\nu}$ is the spacetime metric, and $A_\mu$ is the electromagnetic vector potential. In the semiclassical saddle-point approximation, they obtained
\begin{align}
\mathcal{Z}_6 (\mathcal{Q},T) = \exp\Bigl( -F_{BH}/T \Bigr) \times \Biggl( \quad ....????... \quad \Biggr) \,, \label{b10}
\end{align}
where $F_{BH}$ is the free energy, and $S_{BH}=-\partial F_{BH}/\partial T$ is the Bekenstein-Hawking entropy. In the low $T$ limit at fixed $\mathcal{Q}$, this is
\begin{align}
S_{BH} (T \rightarrow 0, \mathcal{Q}) &= \frac{{\mathcal A}(T) c^3}{4 \hbar G} = \frac{{\mathcal A}_0 c^3}{4 \hbar G}\left(1 + \frac{2 \sqrt{\pi {\mathcal A}_0}\, T}{\hbar c}  + \ldots \right) \,, \label{b1}
\end{align}
where $G$ is Newton's constant, $c$ is the velocity of light, $\hbar$ is Planck's constant, and ${\mathcal A}(T)$ is the area of the charged black hole horizon with
\begin{align}
{\mathcal A}_0 \equiv {\mathcal A}(T \rightarrow 0) = \frac{2 G \mathcal{Q}^2}{c^4} \,. \label{b4}
\end{align}
Thus charged black holes have a non-vanishing zero temperature entropy, a feature known since the earliest days \cite{Gibbons_Hawking}. After the appearance of the SYK model with an extensive non-zero temperature entropy \cite{GPS00,GPS01}, the connection between charged black holes and the SYK model became apparent \cite{SS10} (notice the similarity of the $T$ dependence of (\ref{b1}) to the extensive terms in (\ref{ST2})). The Planckian nature of black hole quasi-normal modes \cite{cvc}, with a relaxation time of order $\hbar/(k_B T)$ shared with the SYK model, was also important for the arguments of Ref.~\cite{SS10} (such quasi-normal modes have been observed \cite{LIGO}).

We now exploit the technology developed for the SYK model to evaluate $\mathcal{Z}_6$ beyond the Gibbons-Hawking saddle-point approximation. Using the dimensional reduction in Fig.~\ref{fig7}, it can be shown that \cite{MSY16,Moitra18,Sachdev19,Iliesiu:2020qvm,GKST}
\begin{align}
& \mathcal{Z}_6 (\mathcal{Q},T)  \nonumber \\
& \approx \exp \left( \frac{{\mathcal A}_0 c^3}{4 \hbar G} \right) \int \mathcal{D} g_{\mu\nu} \mathcal{D} A_\mu \exp \left( - \frac{1}{\hbar} ~ I^{(1+1)}_{\mathrm{JT~gravity~of~AdS}_2+ \mathrm{boundary~graviton}} [g_{\mu\nu}, A_\mu] \right)  \nonumber \\
& = \exp \left( \frac{{\mathcal A}_0 c^3}{4 \hbar G} \right) \int \mathcal{D} f(\tau) \mathcal{D} \phi(\tau) \nonumber \\
&~~~~~\times \exp \left( -\frac{1}{\hbar} I^{0+1}_{{\rm SYK}} [\mbox{time reparameterizations $f(\tau)$, phase rotations $\phi(\tau)$}] \right) \,.  \label{b2}
\end{align}
Here JT gravity is an effective 1+1 dimensional gravity theory obtained from the dimensional reduction, and this describes the dynamics of the boundary graviton in Fig.~\ref{fig7}. Remarkably the JT gravity theory is exactly equivalent to the 0+1 dimensional theory $I^{0+1}_{\rm SYK}$ of time reparameterizations $f(\tau)$ and U(1) phase fluctuations $\phi (\tau)$ obtained from the low energy limit of the SYK action $S_{\rm eff} [ G]$ in (\ref{e5a}). The theory $I^{0+1}_{\rm SYK}$ contains a Schwarzian action for $f(\tau)$.
The final result \cite{Banerjee:2010qc,Sen12,Nayak:2018qej,Moitra18,Sachdev19,Iliesiu:2020qvm,Iliesiu:2022onk,Turiaci_Review} for the black hole density of states is obtained by evaluating the path integral in (\ref{b2}), using (\ref{f5}), and fixing the overall prefactor \cite{Iliesiu:2022onk} by the low-energy physics of the 4-dimensional spacetime 
\begin{align}
D(E,\mathcal{Q}) \sim \left( \frac{{\mathcal A}_0 c^3}{\hbar G} \right)^{-347/90} \exp\left( \frac{{\mathcal A}_0 c^3}{4 \hbar G} \right) \sinh \left( \left[\frac{ \sqrt{\pi} {\mathcal A}_0^{3/2} c^2 }{\hbar^2 G} E \right]^{1/2} \right)\,, \label{b3}
\end{align}
where ${\mathcal A}_0$ is given by (\ref{b4}). Notice the striking similarity to (\ref{de}) for the SYK model. The exponential and sinh factors are in one-to-one correspondence, and in both cases the remaining pre-factor relies upon physics beyond the low energy path integral in (\ref{b2}).

Collecting these results, we also determine the unknown corrections in (\ref{b10}) and (\ref{b1}), and obtain the low $T$ limit of the entropy at fixed $\mathcal{Q}$:
\begin{align}
S (T, \mathcal{Q}) &= \frac{{\mathcal A}_0 c^3}{4 \hbar G}\left(1 + \frac{2 \sqrt{\pi {\mathcal A}_0}\, T}{\hbar c} \right) 
- \frac{3}{2}  \ln \left( \frac{\sqrt{\hbar c^5/G}}{T} \right)  -\frac{559}{180} \ln \left( \frac{{\mathcal A}_0 c^3}{\hbar G} \right) \,. \label{b11}
\end{align}
Again, note the close similarity to (\ref{ST2}) for the SYK model, with the surface area of the black hole in Planck units ${\mathcal A}_0 c^3/(\hbar G)$, and the number of fermionic qubits in the SYK model $N$, playing equivalent roles. Also, the Planck energy $\sqrt{\hbar c^5/G}$ maps to the root-mean-square strength of the SYK interactions $U$. The first 3 terms in (\ref{b11}) are universal in that they apply to any quantum theory of gravity whose low energy limit is non-supersymmetric Einstein-Maxwell theory; a similar universality applies to (\ref{ST2}) for a wide class of SYK-type models. Only the co-efficient on the second logarithm in (\ref{b11}) depends upon more microscopic information \cite{Iliesiu:2022onk,Banerjee:2010qc,Sen12}, as does the co-efficient of the second logarithm in (\ref{ST2}) \cite{GKST}.

Finally, we note that the evaluation of black hole entropy in string theory \cite{Strominger96} relied upon supersymmetry, where the entropy is realized by an {\it exponentially large\/} ground state degeneracy that is present both in supersymmetric SYK models \cite{Fu16} and black holes \cite{Heydeman:2022lse}. Here we have described the generic behavior without low energy supersymmetry, where there is {\it no\/} significant ground state degeneracy. The non-supersymmetric case does have a zero temperature entropy obtained in the limit $T \rightarrow 0$ after $N \rightarrow \infty$, linked to the exponentially small level spacing near the ground state shown in Fig.~\ref{fig6}. The ground state degeneracy is associated with the opposite order of limits (as we noted below (\ref{ST2})), and it is exponentially large only with supersymmetry.

Recent work \cite{Moitra:2019bub,Kapec24a,Kolanowski:2024zrq,Kapec24b} has obtained similar results for the low energy density of states of rotating black holes.

\section{Universal theory of strange metals}
\label{sec:sm}

Understanding the ubiquitous strange metal state of correlated electron materials has been a long-standing challenge in quantum condensed matter physics \cite{Chowdhury:2021qpy,SSORE}. There have been numerous attempts to make the SYK model (\ref{e5}) more realistic by giving it spatial structure by allowing the fermions to hop on a lattice. These attempts have given much insight, but left key issues open: they do give regimes of linear-in-$T$ resistivity, but the low $T$ behavior is inevitably that of a disordered Fermi liquid \cite{Chowdhury:2021qpy}.

Rapid progress has been made recently starting from `Yukawa-SYK' models which randomly couple the fermions to bosons \cite{Fu16,Murugan:2017eto,Patel:2018zpy,Marcus:2018tsr,Wang:2019bpd,Ilya1,Wang:2020dtj,KimAltman20,WangMeng21,Schmalian2,Schmalian3}. The simplest such model has only dispersionless fermions $\psi_i$ and Einstein oscillators $\phi_\ell$ with the same frequency $\omega_0$, coupled to each other with a random Yukawa coupling $g_{i j \ell}$ \cite{Wang:2019bpd,Ilya1}: 
\begin{align}
& H_7 = -\mu\sum_{i} \psi_i^\dagger \psi_i + \sum_{\ell} \frac{1}{2} \left( \pi_\ell^2 + \omega_0^2 \phi_\ell^2 \right) + \sum_{ij\ell} g_{ij\ell} \psi_i^\dagger \psi_j \phi_\ell \nonumber \\
&~~~~ \overline{g_{ij\ell}} = 0, \quad \overline{|g_{ij\ell}|^2} = g^2 , \quad \mbox{different $g_{ij\ell}$ uncorrelated.} \label{e7}
\end{align}
This YSYK model can be solved exactly in the large $N$ limit, and displays a SYK critical state with properties very similar to those of the SYK model in Section~\ref{sec:syk}, including the emergent time reparameterization symmetry. The model $H_7$ also has a superconducting state, but we will not explore that here.

We now wish to make $H_7$ more realistic \cite{Altman1,Patel1,Maria22,Patel2,Guo2022,Schmalian1,PatelLunts,Li:2024kxr,AAPQMC}. As an example, consider the Hertz-Millis theory of quantum phase transition in a metal associated with a density wave order parameter $\phi_a$ ($a = 1 \ldots M$) \cite{VojtaRMP} (similar considerations apply also Fermi-volume changing transitions involving fractionalized bosons \cite{SSORE}). The transition is between a phase with $\langle \phi \rangle \neq 0$ and a phase with $\langle \phi \rangle =0$. In the presence of spatial disorder, we can write the following Lagrangian \cite{Patel14} 
\begin{align}
\mathcal{L}_8 = & \psi_{{\bm k}}^\dagger \left( \frac{\partial}{\partial \tau} + \varepsilon ({\bm k})\right) \psi_{\bm k} +  {\color{blue} v({\bm r})} \psi^\dagger ({\bm r}) \psi  ({\bm r})  + e^{i {\bm G} \cdot {\bm r}} g \, \psi^{\dagger}({\bm r}) \,\mathcal{M}\,  \psi({\bm r}) \, \phi({\bm r}) \nonumber \\
&~~~~~~~~+ K \, [ \nabla_{\bm r} \phi ({\bm r})]^2 + [s+ {\color{blue} \delta s ({\bm r})}] \,[\phi({\bm r})]^2+u  \,[\phi({\bm r})]^4 \,, \label{e8}
\end{align}
where the couplings with a fixed random dependence on space are shown in {\color{blue} blue}.
In comparison to $H_7$, in $\mathcal{L}_8$ we have given the fermions a Fermi surface associated with the dispersion $\varepsilon ({\bm k})$, and a spatial field theory for the scalar field $\phi$ in $d=2$ spatial dimensions. The coupling matrix $\mathcal{M}$ and the ordering wavevector ${\bm G}$ are fixed, and their form depends upon the particular symmetry being broken. The transition is tuned by varying the boson `mass' $s$. Crucial to our considerations are the ${\bm r}$ dependent couplings which represent the spatial disorder; these obey the disorder averages
\begin{itemize}
\item
Spatially random potential ${\color{blue} v ({\bm r})}$ with ${\color{blue} \overline{v({\bm r})} = 0}$, ${\color{blue} \overline{v({\bm r}) v({\bm r'})} = v^2 \delta({\bm r}-{\bm r'})}$.
\item
Spatially random mass ${\color{blue} \delta s ({\bm r})}$ with ${\color{blue} \overline{\delta s ({\bm r})} = 0}$, ${\color{blue} \overline{\delta s ({\bm r}) \delta s ({\bm r'})} = \delta s^2  \delta({\bm r}-{\bm r'})}$.
\end{itemize}

Without spatial randomness in $\mathcal{L}_8$, we obtain a Fermi surface with breakdown of quasiparticles at `hot spots', but nevertheless, with perfect metal behavior in transport with zero resistivity \cite{Patel14}.
A standard analysis of $\mathcal{L}_8$ for weak disorder shows that the random mass disorder ${\color{blue} \delta s ({\bm r})}$ is the strongest relevant perturbation at the quantum phase transition \cite{Patel14}. This is a consequence of the violation of the Harris criterion $\nu > 2/d$ \cite{CCFS}, and the value $\nu =1/2$ in $d=2$. 

Inspired by various works \cite{Chowdhury:2021qpy} on making the SYK model more realistic, recent work \cite{Patel2} proposed the following approach for dealing with this strongly relevant perturbation, to obtain a theory which can be treated in a SYK-like disorder-self-averaging.
We can account for the spatial dependence of ${\color{blue} \delta s ({\bm r})}$ by expressing the $\phi$ quantum fluctuations in a new basis which diagonalizes the $\phi$ quadratic form. As long as the $\phi$ eigenstates remain extended, we expect a diffusive form for the $\phi$ propagator \cite{HLR}. But the transformation to this new basis induces disorder in the Yukawa coupling, as in (\ref{e7}), leading to a 2dYSYK theory
\begin{align}
\mathcal{L}_9 = & \psi_{{\bm k}}^\dagger \left( \frac{\partial}{\partial \tau} + \varepsilon ({\bm k})\right) \psi_{\bm k}  +s \,[\phi({\bm r})]^2 + e^{i {\bm G} \cdot {\bm r}} [g +  {\color{blue} g'({\bm r})}] \, \psi^{\dagger}({\bm r}) \,\mathcal{M}\,  \psi({\bm r}) \, \phi({\bm r}) \nonumber \\
&~~~~~~~~+ K \, [ \nabla_{\bm r} \phi ({\bm r})]^2 +u  \,[\phi({\bm r})]^4 +  {\color{blue} v({\bm r})} \psi^\dagger ({\bm r}) \psi  ({\bm r})\,, \label{e9}
\end{align}
with the coupling ${\color{blue}  g' ({\bm r})}$ obeying 
\begin{itemize}
\item Spatially random Yukawa coupling ${\color{blue} g' ({\bm r})}$ with ${\color{blue} \overline{g'({\bm r})} = 0}$,\\ ${\color{blue} \overline{g'({\bm r}) g'({\bm r'})} = g'^2 \delta({\bm r}-{\bm r'})}$.
\end{itemize}
The transformation to the new $\phi$ basis will also induce spatial disorder in other couplings in (\ref{e9}), but we ignore these less important terms. We further assume that in the regime of extended $\phi$ eigenstates we can treat the ${\color{blue} g' ({\bm r})}$ and ${\color{blue} v({\bm r})}$ disorder in a self-averaging manner, similar to that in the YSYK model (this becomes exact after adding a large number of flavor indices to (\ref{e9}), as in (\ref{e7})).  
Then we obtain self-consistent equations for the boson and fermion Green's functions analogous to (\ref{sy}) for the SYK model \cite{Li:2024kxr}. Because of the spatial dimensionality $d=2$, these equations involve Green's functions with momentum and position arguments. The SYK method can also be extended to compute transport properties of $\mathcal{L}_9$ \cite{Guo2022,Li:2024kxr}. The results from such computations \cite{Li:2024kxr} agree with numerous observations on strange metals at intermediate temperatures, including the $T$-dependent resistivity, the optical conductivity \cite{Michon23}, the specific heat, and the marginal Fermi liquid behavior of the electron spectral function \cite{Reber2019}.

In recent work \cite{PatelLunts,AAPQMC}, the proposed equivalence between $\mathcal{L}_8$ and $\mathcal{L}_9$ was examined more carefully by studying the spatial structure of the $\phi$ eigenmodes. 
Computation of Altshuler-Aronov corrections to the $d=2$ SYK theory with only ${\color{blue} v ({\bm r})}$ disorder show singular corrections requiring reconsideration of the $\phi$ propagator \cite{Foster22,Grilli22,Grilli23}. We interpreted these corrections as a signal of the strong relevance of ${\color{blue} \delta s ({\bm r})}$ in $\mathcal{L}_8$ or of {\color{blue} $g' ({\bm r})$} in $\mathcal{L}_9$ at very low energies: we can expect the $\phi$ eigenmodes to localize, and then ${\color{blue} g' ({\bm r})}$ cannot be treated in a self-averaging manner \cite{PatelLunts}. From an application of the strong disorder renormalization group to large ${\color{blue} |\delta s ({\bm r})|}$, Hoyos {\it et al.\/} \cite{Hoyos07}
had proposed that the lowest energy bosonic eigenmodes have a structure very similar to that of the random quantum Ising model, specifically to the extension of the model $H_2$ in (\ref{e2}) to the ferromagnetic case with a non-zero $J_{ij}$ average in spatial dimension $d=2$ \cite{Motrunich00}. 
Ref.~\cite{PatelLunts} examined the bosonic eigenmodes of $\mathcal{L}_8$ by integrating out the fermions $\psi$, and numerically diagonalizing  the $\phi$ quadratic form in a Hartree approximation \cite{Adrian08}. 
\begin{figure}
\begin{center}
\includegraphics[width=4.5in]{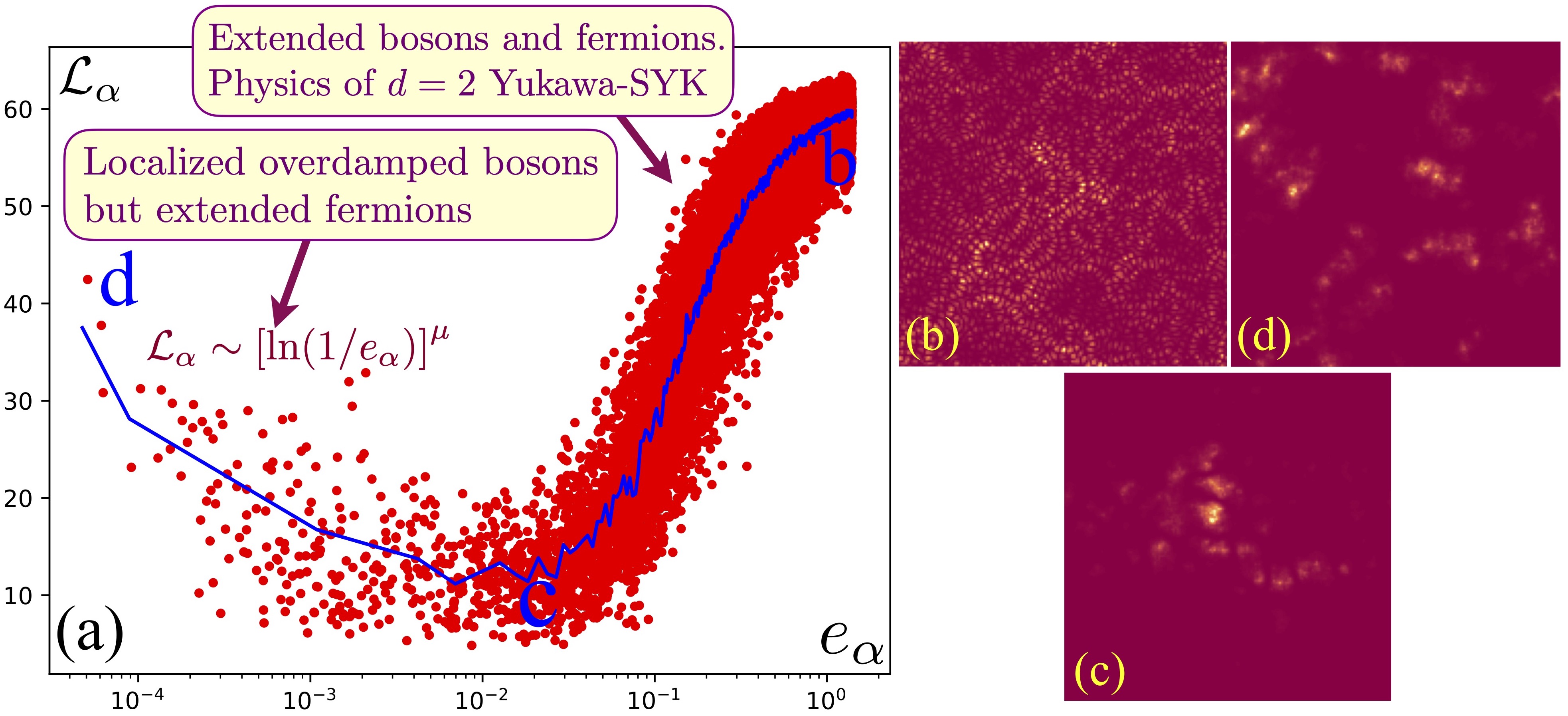}
\end{center}
\caption{Adapted from Ref.~\cite{PatelLunts}. (a) Localization length $\mathcal{L}_\alpha$ of overdamped bosonic eigenmodes of (\ref{e8}) as a function of their energy $e_\alpha$, computed in a large $M$ limit. (b,c,d) Pictures of the corresponding bosonic eignfunctions. The self-averaging SYK-like theory of $\mathcal{L}_9$ applies to (b), while the mapping to the random Ising model applies to (d).}
\label{fig8}
\end{figure}
Fig.~\ref{fig8} shows numerical results for the localization length of the bosonic eigenmodes of $\mathcal{L}_8$ from Ref.~\cite{PatelLunts}. Similar results appear in the quantum Monte Carlo studies of $\mathcal{L}_9$ \cite{AAPQMC}, and this is strong evidence for the equivalence of $\mathcal{L}_8$ and $\mathcal{L}_9$. The non-monotonic behavior as a function of energy in Fig.~\ref{fig8}a clearly demonstrates two distinct regimes:
\begin{itemize} 
\item At higher energies, we have extended states, for which the application of self-averaging SYK methods to $\mathcal{L}_9$ works well. 
\item At very low energy, we obtain strong disorder dominated overdamped bosonic modes, whose localization length increases logarithmically slowly with decreasing energy---this is just as expected from the random quantum Ising model \cite{Motrunich00}. Self-averaging methods cannot be applied in this regime.
\end{itemize}
The results of the analyses described above capture the main features of the global phase diagram of strange metals \cite{PatelLunts}, including the `foot' feature which extends strange metal behavior at low $T$ beyond the usual quantum critical region \cite{Hussey_foot}.

\section*{Acknowledgments}

I thank Aavishkar Patel for a productive long-term collaboration \cite{Patel14,Patelshear,Patelchaos,PatelArovas,Patel:2018zpy,PatelPlanckian,Patel1,Maria22,Patel2,Guo2022,Patelshotnoise23,Schmalian1,PatelLunts,Li:2024kxr,Lunts24} on the topics in Section~\ref{sec:sm}. I also thank Joaquin Turiaci and Sameer Murthy for valuable comments on black holes. This research was supported by the U.S. National Science Foundation grant No. DMR-2245246, and by the Simons Collaboration on Ultra-Quantum Matter which is a grant from the Simons Foundation (651440, S.S.). 

\bibliographystyle{JHEP}
\bibliography{SachdevSolvay29}

\end{document}